\documentclass[aps,preprintnumbers,amsmath,amssymb,nofootinbib]{revtex4}
\usepackage{graphicx}

\begin{document}

\preprint{gr-qc/0309069}

\title{
Quark stars admitting a one-parameter group of conformal motions}

\author{M. K. Mak}
\email{mkmak@vtc.edu.hk} \affiliation{ Department of Physics, The
University of Hong Kong, Pokfulam Road, Hong Kong}

\author{T. Harko}
\email{harko@hkucc.hku.hk} \affiliation{ Department of Physics,
The University of Hong Kong, Pokfulam Road, Hong Kong}

\date{January 13, 2004}


\begin{abstract}

An exact analytical solution describing the interior of a charged
strange quark star is found under the assumption of spherical
symmetry and the existence of a one-parameter group of conformal
motions. The solution describes a unique static charged
configuration of quark matter with radius $R=9.46$ km and total
mass $M=2.86M_{\odot }$.

\end{abstract}


\maketitle

\section{Introduction}

It is largely believed today that the strange quark matter, consisting of
the u,d and s quarks is the most energetically favorable state of baryon
matter. Witten \cite{Wi} specified the two ways of formation of the strange
matter: the quark-hadron phase transition in the early universe and
conversion of neutron stars into strange ones at ultrahigh densities. Quark
bag models in the theories of strong interactions suppose that the breaking
of physical vacuum takes place inside hadrons. As a result the vacuum energy
densities inside and outside a hadron become essentially different and the
vacuum pressure $B$ on a bag wall equilibrates the pressure of quarks thus
stabilizing the system.

The structure of a realistic strange star is quite complicated and can be
described as follow \cite{Ch}. Beta-equilibrated strange quark - star matter
consists of an approximately equal mixture of up, down and strange quarks,
with a slight deficit of the latter. The Fermi gas of $3A$ quarks constitutes
a single color-singlet baryon with baryon number A. This structure of the
quarks leads to a net positive charge inside the star. Since stars in their
lowest energy state are supposed to be charge neutral, electrons must
balance the net positive quark charge in strange matter stars. Being bounded
by the Coulomb force, rather than the strong one, as is the case for quarks,
the electrons extend several hundred fermis beyond the surface of the
strange star. Associated with this electron displacement is a very strong
electric dipole layer that can support, out of contact with the surface of a
strange star, a crust of nuclear material, which it polarizes. The neutron
drip density determines the maximal possible density at the base of the
crust (the inner crust density) \cite{Ch}. Being electrically charge neutral the
neutrons do not feel the Coulomb force and hence would gravitate toward the
quark core where they become converted into strange quark matter.

The theory of the equation of state of strange stars is directly based on
the fundamental QCD Lagrangian \cite{We}
\begin{equation}
L_{QCD}=\frac{1}{4}\sum_{a}F_{\mu \nu }^{a}F^{a\mu \nu }+\sum_{f=1}^{N_{f}}%
\bar{\psi}\left( i\gamma ^{\mu }\partial _{\mu }-g\gamma ^{\mu }A_{\mu }^{a}%
\frac{\lambda ^{a}}{2}-m_{f}\right) \psi ,
\end{equation}
where the subscript $f$ denotes the various quark flavors $u$, $d$, $s$, $c$
etc. and the nonlinear gluon field strength is given by
\begin{equation}
F_{\mu \nu }^{a}=\partial _{\mu }A_{\nu }^{a}-\partial _{\nu }A_{\mu
}^{a}+gf_{abc}A_{\mu }^{b}A_{\nu }^{c}.
\end{equation}

QCD predicts a weakening of the quark-quark interaction at short distances

Assuming that interactions of quarks and gluons are sufficiently small the
energy density $\varepsilon $ and pressure $p$ of a quark -gluon plasma at
temperature $T$ and chemical potential $\mu _{f\text{ }}$can be calculated
by thermal theory. Neglecting quark masses in first order perturbation
theory, the equation of state is given by \cite{Wi}, \cite{Ch}:
\begin{equation}
\varepsilon =\sum_{i=u,d,s,c;e^{-},\mu ^{-}}\varepsilon
_{i}+B,p+B=\sum_{i=u,d,s,c;e^{-},\mu ^{-}}p_{i},
\end{equation}
where $B$ is the difference between the energy density of the perturbative
and non-perturbative QCD vacuum (the bag constant) . Hence the pressure -
energy density relation for quark matter is given by the MIT bag model
equation of state \cite{Wi}, \cite{Ch}
\begin{equation}
p=\frac{1}{3}\left( \rho -4B\right) .  \label{state}
\end{equation}

Equation (\ref{state}) is essentially the equation of state of a gas of
massless particles with corrections due to the QCD trace anomaly and
perturbative interactions. These are always negative, reducing the energy
density at given temperature by about a factor two when $\alpha _{s}=0.5$
\cite{We}.

Most of the investigations of the stellar quark-gluon plasma have
been done under the assumption of the electric charge neutrality of the
quark-gluon plasma that reads $\sum_{i=u,d,s,c;e^{-},\mu ^{-}}q_{i}n_{i}=0$.
In the case of a star formed from massless u, d and s quarks the charge
neutrality condition can be explicitly formulated as $2n_{u}/3=\left(
n_{d}+n_{s}\right) /3$ \cite{Ch}.

There are several proposed mechanisms for the formation of quark
stars. Quark stars are expected to form during the collapse of the core of a
massive star after the supernova explosion as a result of a first or second
order phase transition, resulting in deconfined quark matter \cite{Da}. The
proto-neutron star core or the neutron star core is a favorable environment
for the conversion of ordinary matter to strange quark matter \cite{Da}. Another
possibility is that some neutron stars in low-mass X-ray binaries can
accrete sufficient mass to undergo a phase transition to become strange
stars \cite{Ch2}. This mechanism has also been proposed as source of radiation
emission for cosmological $\gamma $-ray bursts \cite{Ch3}.

It is the purpose of the present Letter to consider the behavior
of strange quark stars described by the MIT bag model with respect
to one class of admissible transformations: conformal motions. We
shall also explore the physical consequences on the structure and
properties of the strange stars of the existence of this
transformation.

As a group of admissible transformations for a strange star we
shall consider spherically symmetric and static distributions of quark
matter that admits a one-parameter group of conformal motions, i.e.
\begin{equation}
L_{\xi }g_{ik}=\phi \left( r\right) g_{ik},  \label{lie}
\end{equation}
where the left-hand side is the Lie derivative of the metric tensor
describing the interior gravitational field of a strange star with respect
to the vector field $\xi ^{i}$ and $\phi \left( r\right) $ is an arbitrary
function of the radial coordinate r. This type of symmetry has been
intensively used to describe the interior of neutral or charged
general relativistic stellar-type objects \cite{He}, \cite{He1}, \cite{He2}. With the assumption (\ref{lie}) the
gravitational field equations describing the interior of a strange star can
be integrated in Schwarzschild coordinates and an exact simple physical
solution, corresponding to a charged strange star, can be obtained.

The present paper is organized as follows. In Section II we obtain
the general solution for the Einstein-Maxwell equations for a
static charged configuration with the matter content obeying the
bag model equation of state under the assumption that the
space-time admits a one-parameter group of conformal motions. In
Section III we discuss and conclude our results.

\section{Strange quark stars admitting conformal symmetry}

In the following we shall adopt geometrized units such that $8\pi G=c=1$ .
The sign conventions used are those of the Landau-Lifshitz timelike
convention. Let us consider a spherically symmetric static distribution of
strange quark matter. In Schwarzschild coordinates the line element takes
the following form:
\begin{equation}
ds^{2}=e^{\nu \left( r\right) }dt^{2}-e^{\lambda \left( r\right)
}dr^{2}-r^{2}\left( d\theta ^{2}+\sin ^{2}\theta d\chi ^{2}\right) .
\label{9}
\end{equation}

The total energy-momentum tensor $T_{i}^{k}$ inside the strange star is
assumed to be the sum of two parts $M_{i}^{k}$ for the quark matter and $%
E_{i}^{k}$ for an electromagnetic contribution, respectively: $%
T_{i}^{k}=M_{i}^{k}+E_{i}^{k}$.

The energy-momentum tensor for the quark matter has the usual expression
\begin{equation}
M_{i}^{k}=\left( \rho +p\right) u_{i}u^{k}-p\delta _{i}^{k},  \label{11}
\end{equation}
with $p$ and $\rho $ related by the bag model equation of state  (\ref{state}%
). In equation (\ref{11}) $u^{i}=\delta _{0}^{i}e^{-\nu /2}$ is the
four-velocity satisfying $u_{i}u^{i}=1$.

The electromagnetic contribution is given by\qquad \qquad \qquad
\begin{equation}
E_{i}^{k}=-\frac{1}{4\pi }\left( F_{il}F^{kl}-\frac{1}{4}\delta
_{i}^{k}F_{lm}F^{lm}\right) ,
\end{equation}
where $F_{ik}$ is the electromagnetic field tensor defined in terms
of the four-potential $A_{i}$ as
\begin{equation}
F_{ik}=A_{i,k}-A_{k,i}.
\end{equation}

For the electromagnetic field we shall adopt the gauge $A_{i}=\left( \varphi
\left( r\right) ,0,0,0\right) $.

The Einstein-Maxwell equations describing the interior of a charged strange
star can be expressed as
\begin{equation}
R_{i}^{k}-\frac{1}{2}\delta
_{i}^{k}R=T_{i}^{k},F_{ik,l}+F_{li,k}+F_{kl,i}=0,F_{;k}^{ik}=-j^{i}/2,
\label{field}
\end{equation}
where $j^{i}=\bar{\rho}_{e}u^{i}$ is the four-current density and $\bar{\rho}%
_{e}$ is the proper charge density.

Using the line element (\ref{9}) the field equations (\ref{field}) take the
form (we denote the derivative with respect to $r$ by a prime):
\begin{equation}
\rho +E^{2}=-e^{-\lambda }\left( \frac{1}{r^{2}}-\frac{\lambda ^{^{\prime }}%
}{r}\right) +\frac{1}{r^{2}},  \label{field1}
\end{equation}
\begin{equation}
-p+E^{2}=-e^{-\lambda }\left( \frac{\nu ^{^{\prime }}}{r}+\frac{1}{r^{2}}%
\right) +\frac{1}{r^{2}},
\end{equation}
\begin{equation}
p+E^{2}=\frac{1}{2}e^{-\lambda }\left( \nu ^{^{\prime \prime }}+\frac{\left(
\nu ^{^{\prime }}\right) ^{2}}{2}+\frac{\nu ^{^{\prime }}-\lambda ^{^{\prime
}}}{r}-\frac{\nu ^{^{\prime }}\lambda ^{^{\prime }}}{2}\right) ,
\end{equation}
\begin{equation}
\frac{d}{dr}\left( r^{2}E\right) =\frac{1}{2}\rho _{e}r^{2}.  \label{field4}
\end{equation}

In equations (\ref{field1})-(\ref{field4}) $E$ is the usual electric field
intensity defined as $E^{2}=-F_{01}F^{01}$ and $E\left( r\right) =e^{-\left(
\nu +\lambda \right) /2}\varphi ^{^{\prime }}\left( r\right) $ with $\varphi
^{^{\prime }}\left( r\right) $ =$F_{10}$. The charge density $\rho _{e}$ in
equation (\ref{field4}) is related to the proper charge density $\bar{\rho}%
_{e\text{ }}$by $\rho _{e}=\bar{\rho}_{e\text{ }}e^{\lambda /2}$. We first
integrate equation (\ref{field4}) to obtain $E\left( r\right) =Q\left(
r\right) /r^{2}$, where $Q\left( r\right) =\frac{1}{2}\int_{0}^{r}\rho
_{e}r^{2}dr=\frac{1}{2}\int_{0}^{r}\bar{\rho}_{e\text{ }}e^{\lambda
/2}r^{2}dr$ is the charge within radius $r$.

Now we shall assume that the space-time describing the interior of
the strange quark star admits a one-parameter group of conformal motions (%
\ref{lie}), i.e.
\begin{equation}
L_{\xi }g_{ik}=\xi _{i;k}+\xi _{k;i}=\phi \left( r\right) g_{ik},
\label{conf}
\end{equation}
with $\phi \left( r\right) $ an arbitrary function of $r$.

Using the line element (\ref{9}) equation (\ref{conf}) explicitly
reads
\begin{equation}
\xi ^{i}\nu ^{^{\prime }}=\phi ,\xi ^{0}=\bar{C}=const.,\xi ^{1}=\phi
r/2,\lambda ^{^{\prime }}\xi ^{1}+2\frac{d\xi ^{1}}{dr}=\phi .  \label{conf1}
\end{equation}

Equations (\ref{conf1}) have the general solution given by \cite{He}
\begin{equation}
e^{\nu }=A^{2}r^{2},\phi =Ce^{-\lambda /2},\xi ^{i}=\bar{C}\delta
_{0}^{i}+\phi r\delta _{1}^{i}/2,  \label{conf2}
\end{equation}
with $A$ and $C$ arbitrary constant of integration.

Hence the requirement of the existence of conformal motions imposes strong
constraints on the form of the metric tensor coefficients for a strange
star. Substituting (\ref{conf2}) into the field equations (\ref{field1})-(%
\ref{field4}) we obtain
\begin{equation}
\rho +E^{2}=\frac{1}{r^{2}}\left( 1-\frac{\phi ^{2}}{C^{2}}\right) -\frac{2}{%
C^{2}}\frac{\phi \phi ^{^{\prime }}}{r},-p+E^{2}=\frac{1}{r^{2}}\left( 1-3%
\frac{\phi ^{2}}{C^{2}}\right) ,p+E^{2}=\frac{1}{C^{2}}\frac{\phi ^{2}}{r^{2}%
}+\frac{2}{C^{2}}\frac{\phi \phi ^{^{\prime }}}{r}.  \label{20}
\end{equation}

We can formally solve the field equations (\ref{20}) and express $%
p,\rho $ and $E^{2}$ as
\begin{equation}
\rho =-\frac{3Y^{^{\prime }}}{2r}+\frac{1}{2r^{2}},p=\frac{Y^{^{\prime }}}{2r%
}+\frac{4Y-1}{2r^{2}},E^{2}=\frac{Y^{^{\prime }}}{2r}+\frac{1-2Y}{2r^{2}},
\label{21}
\end{equation}
where we have introduced a new variable $Y=\frac{\phi ^{2}}{C^{2}}$.

From equations (\ref{21}) we obtain, by using the bag equation of state of
the strange matter (\ref{state}), the following differential equation for $Y$%
:
\begin{equation}
\frac{dY}{dr}=-\frac{2Y}{r}+\frac{2}{3r}-\frac{4B}{3}r.  \label{equ}
\end{equation}

Eq. (\ref{equ}) has the general solution given by
\begin{equation}
Y\left( r\right) =\frac{1}{r^{2}}\left[ C_{1}+\frac{r^{2}}{3}\left(
1-Br^{2}\right) \right],
\end{equation}
with $C_{1}$ an arbitrary integration constant.

In order to obtain finite and well-defined values of the energy
density and mass for all $r\leq R$ (we denote by $R$ the radius of the
strange star) it is necessary that the arbitrary integration constant $C_{1}$
be zero. Therefore we obtain the following explicit exact solution describing the
interior of a charged strange star:
\begin{equation}
e^{\nu }=A^{2}r^{2},e^{\lambda }=\frac{3}{1-Br^{2}},\rho =\frac{1}{2r^{2}}%
+B,p=\frac{1}{6r^{2}}-B,E^{2}=\frac{1}{6r^{2}}.  \label{24}
\end{equation}

The radius $R$ of the static strange quark matter configuration can be
obtained from the condition of the vanishing pressure at the surface of the
star, $p\left( R\right) =0$ and is given by $R=1/\sqrt{6B}$. Using the
results $E=Q\left( r\right) /r^{2}$, $E^{2}=1/\left( 6r^{2}\right) $ and $%
R=1/\sqrt{6B}$, we obtain the charge distribution in the interior of the
quark star in the form $Q=r/\sqrt{6}=\sqrt{B}Rr$. The charge is linearly
increasing with the radius $r$.

The solution of the Einstein-Maxwell equations for $r>R$ is given by the
Reissner-Nordstrom metric as
\begin{equation}
ds^{2}=\left( 1-2M/r+Q^{2}/r^{2}\right) dt^{2}-\left(
1-2M/r+Q^{2}/r^{2}\right) ^{-1}dr^{2}-r^{2}\left( d\theta ^{2}+\sin
^{2}\theta d\chi ^{2}\right) ,
\end{equation}
where $M$ and $Q$ are the total mass and charge of the strange star
respectively. To match the line element (\ref{9}) with the
Reissner-Nordstrom metric across the boundary at $r=R$ we require the
continuity of the gravitational potentials and of the radial electric field
at $r=R$ . The continuity of $e^{\nu }$ leads to the determination of the
constant $A$ as $A^{2}=\left( 1-2M/R+Q^{2}/R^{2}\right) /R^{2}$, while the
continuity of the electric field leads to a total charge-radius relation of
the form $Q\left( R\right) =R/\sqrt{6}=\sqrt{B}R^{2}$. The intensity of the
electric field on the surface of the strange star is given by $E\left(
R\right) =Q\left( R\right) /R^{2}=\sqrt{B}$. The continuity of on the
boundary $r=R$ leads to a mass charge relation of the form $M=2Q^{2}\left(
R\right) /\left( 3R\right) +R/3$. On the other hand the total mass of the
quark matter component of the star is given by
\begin{equation}
M_{q}=\frac{1}{2}\int_{0}^{R}\rho r^{2}dr=BR^{3}/6+R/4.
\end{equation}

The distribution of the proper charges density inside the quark star follows
from equations (\ref{field1}) and (\ref{field4}) and is given by $\bar{\rho}%
_{e\text{ }}=\sqrt{2\left( 1-Br^{2}\right) }/\left( 3r^{2}\right)$.
Immediately, by taking $r=R=1/\sqrt{6B}$, the proper surface charge density
is given by $\bar{\rho}_{e\text{ }}\left( R\right) =\sqrt{\frac{20}{3}}B$.

Hence we have obtained the complete solution of the gravitational
field equations for a charged strange quark star described by the MIT bag
model.

\section{Discussions and final remarks}

A physically acceptable interior solution of the gravitational field
equations must comply with the following conditions: a) the matter density $%
\varepsilon $ and the fluid pressure p should be non-negative throughout the
distribution b) the gradients $d\rho /dr$ and $dp/dr$ should be negative c)
the speed of sound should not exceed the speed of light as implication of
causality fulfillment d) the interior metric should match continuously with
an exterior solution.

The solution to the gravitational field given by
equations (\ref{24}) satisfies all these four criteria, since we have, $%
\rho \geq 0$, $p\geq 0$, $d\rho /dr$ $=-1/r^{3}<0$ and $dp/dr$ $%
=-1/\left( 3r^{3}\right) <0$ for all $0\leq r\leq R$. The speed of
sound is given by $v_{s}=c/\sqrt{3}$. The singularity in the charge density
and mass density is physically acceptable since the total charge and mass are
finite.

The strange star model with conformal motions describes a single stable quark matter configuration
with radius given by $R=c/\sqrt{48\pi GB}=9.46km$ and with the mass of the
quark matter $M_{q}=4\pi BR^{2}/3+4\pi Rc^{2}/\left( 16\pi G\right) =40\pi
BR^{3}/3\approx 3.545\times 10^{33}g\approx 1.772M_{\odot }$, where $%
B=10^{14}g/cm^{3}$ and $M_{\odot }=2\times 10^{33}g$ have been used. The
total mass of the star (including the electromagnetic contribution) is $%
M=2.86M_{\odot }$.

A complete description of static strange stars has been obtained
based on numerical integration of mass continuity and TOV (hydrostatic
equilibrium) equations for different values of the bag constant. Using numerical methods the maximum gravitational mass $M_{max}$%
, the maximum baryon mass $M_{B,max}\equiv 1.66\times 10^{-27}kg\times N_{B}$
($N_{B}$-the total baryon number of the stellar configuration) and the
maximum radius $R_{max}$ of the strange star , have been obtained, as a
function of the bag constant, in the form \cite{Wi},\cite{Al}, \cite{Ha}, \cite{Ha89}, \cite{Go}:
\begin{equation}
M_{max}=\frac{1.9638M_{\odot }}{\sqrt{B_{60}}},M_{B,max}=\frac{%
2.6252M_{\odot }}{\sqrt{B_{60}}},R_{max}=\frac{10.172km}{\sqrt{B_{60}}},
\end{equation}
where $B_{60}\equiv B/(60MeVfm^{-3})$.

Hence the presence of charge leads to a considerable increase in the
mass of a stellar object obeying the MIT bag model equation of state.
On the other hand the radius of the charged configuration is smaller than
the maximum radius of the uncharged strange star.

The requirement that the strange quark star obeying the MIT equation
of state admits the group of conformal motions is the simplest way that
leads to an exact quark star model. But this requirement also necessities,
in order to close the gravitational field equations, the introduction of an
extra term in the energy-momentum tensor. In order to obtain a complete
physical solution, especially one satisfying condition d) of physical
acceptability in the present Letter we have supposed that the extra-term in
the energy-momentum tensor is due to the presence of a static charge
distribution inside the strange quark star. Other choices are also possible
by adding to the energy-momentum tensor of the perfect quark fluid extra
terms that corresponds to a magnetic field, possible anisotropic stresses
related to some corrections to the MIT bag model state equation,
super-fluidity etc. The arbitrariness in adding the extra term to the energy
-momentum tensor also leads to the possibility of minimizing the
contribution of the non-perfect fluid term. On the other hand a linearly
increasing charge density resulting in a highly charged strange star surface
is not excluded by realistic quark star models \cite{Ch}.

If $B=0$ the solution (\ref{24}) is identical with the solution to
the gravitational field equations obtained by Misner and Zapolsky,
describing neutral matter obeying a $\gamma $-law equation of state \cite{Mi}.

Whether the strange star model presented in this paper actually
describes a well-determined stellar structure can only be decided
once reliable knowledge about the mechanisms that characterize
strange and neutron star formation becomes available from the
underlying theory of stellar evolution and from observational
data.

\end{document}